\documentstyle[12pt,aaspp4]{article}

\begin{document}

\title{General relativistic effects in the neutrino-driven
	wind and $r$-process nucleosynthesis}
\author{Christian Y. Cardall and George M. Fuller}
\affil{Department of Physics, University of 
	California, San Diego,
	La Jolla, CA, 92093-0319}

\begin{abstract}
We discuss general relativistic effects in the steady-state
neutrino-driven ``wind'' which may arise from nascent neutron
stars. In particular,  
we generalize previous analytic estimates of the entropy per
baryon $S$, the mass outflow rate $\dot M$, and the dynamical
expansion time scale $\tau_{\rm dyn}$. 
We show that $S$ increases and 
$\tau_{\rm dyn}$ decreases with increasing values of the 
mass-to-radius ratio describing the supernova core. Both
of these trends indicate that a more compact core will
lead to a higher number of neutrons per iron peak seed
nucleus. Such an enhancement in the neutron-to-seed ratio
may be required for successful $r$-process nucleosynthesis
in neutrino-heated supernova ejecta.
\end{abstract}


The production site of the $r$-process elements 
(Burbidge et al. 1957; Cameron 1957) is a longstanding
problem (Mathews \& Cowan 1990). 
One of the most promising candidate sites for $r$-process
nucleosynthesis is the 
neutrino-heated ejecta from the post--core-bounce 
environment of a
Type II or Type Ib supernova (Meyer et al. 1992; 
Woosley \& Hoffman
1992; Takahashi, Witti, \& Janka 1994; Woosley et al. 1994).
These $r$-process calculations, though promising, 
cannot reproduce a solar system $r$-process abundance
pattern without an artificial increase in the neutron-to-seed
nucleus ratio over that predicted in hydrodynamical calculations
and simple wind models (see, e.g., Hoffman, Qian, \& 
Woosley 1996;
Meyer, Brown, \& Luo 1996).

Qian \& Woosley (1996, hereafter QW) used a simple model
of the neutrino-driven wind to obtain both analytic 
and numerical estimates of 
quantities upon which the nucleosynthesis abundance yield 
depends. 
These quantities
include the electron fraction $Y_e$, the entropy per
baryon $S$, the mass outflow rate $\dot M$, and the 
dynamic expansion
time scale $\tau_{\rm dyn}$---all of which are 
important in setting
the neutron-to-seed nucleus ratio prior to the epoch of rapid
neutron capture.
 
In this {\em Letter} we generalize the analytic derivations 
in QW to
include general relativistic effects. Such effects are 
of potential
interest in light of ``soft'' nuclear matter equations of 
state---involving, for example, kaon 
condensation (Thorsson, Prakash,
\& Lattimer 1994)---which could 
lead to very compact supernova cores
and even black holes as supernova remnants (Bethe \& Brown 1995; 
Woosley \& Timmes 1996).
While QW reported two sample numerical
calculations involving post-Newtonian corrections, they did not
present corresponding analytic calculations. 
Numerical calculations reported in QW suggest that 
general relativistic
effects go in the direction of making conditions more 
favorable for
the $r$-process. We here allow for general relativistic 
effects in semi-analytical calculations, 
including effects not included in the 
numerical calculations by QW,
namely the redshift of neutrino energies and bending of the
neutrino trajectories. 

In the analytic approximations performed by QW,
the calculation of the quantities $S$, $\dot M$, 
and $\tau_{\rm dyn}$
essentially decouples from the calculation of $Y_e$. We will
present semi-analytic estimates for $S$, $\dot M$, and
$\tau_{\rm dyn}$; general relativistic effects on 
$Y_e$ have been considered in another paper 
(Fuller \& Qian 1996).
Unless units are explicitly given, 
we take $\hbar = c = k = G = 1$.

The wind equations in Duncan, Shapiro, and Wasserman (1986) 
and QW can be generalized to allow for relativistic 
outflow velocities
and general relativistic effects in a static 
Schwarzschild spacetime. 
A detailed derivation and discussion
will appear elsewhere (Cardall \& Fuller 1997, hereafter 
CF). We here simply
present the general relativistic analogs of 
Eqs. (24-26) of QW, which
give the radial evolution of the velocity, 
``flow energy'' per baryon
$\epsilon_{\rm flow}$, and entopy per baryon:
\begin{eqnarray}
{v\over {1-v^2}}\!\! \left(1 - 
	{v_s^2\over v^2}\right)\!\! {dv\over dr}
	\!\!\! &=&\!\!\! 
	{1\over r}\left[{2\over 3}
	{TS\over m_N} - {{(1 - TS/3m_N)\over(1 - 2M/r)}}
	{M\over r}\right] - {q\over 3v {y(1 + TS/m_N)}},  \\
& & \nonumber \\
{q \over {(1 + TS/m_N)}}\!\!\!&=&\!\!\! v{y} {d\over dr}\left[
	{\ln\left(1+{{TS\over m_N}}\right)}
	- {1\over 2}{\ln\left(1-{v^2}\right)} +
	 {{1\over 2}\ln\left(1-{ 2 M\over r}\right)}
	\right] \!\!\equiv \!\! vy {d\over dr}
	\epsilon_{\rm flow}, \label{eflow}\\
& & \nonumber \\
v{y} {dS \over dr}\!\!\! &=&\!\!\! 
	{m_N\, q \over T}. \label{entropy1}
\end{eqnarray}
In these equations, $v$ is the outflow velocity as measured
by an observer at rest in the Schwarzschild spacetime;
$v_s \equiv (TS/3m_N)^{1/2}$ is the sound speed; $T$ is the
temperature; $S$ is the entropy per baryon; $m_N$ is the
baryon rest mass; $M$ is the mass of the supernova core; 
$q$ is the heating rate per unit mass; and $y \equiv 
\left[(1-2M/r)/(1-v^2)\right]^{1/2}$. 
These equations assume radiation dominated conditions,
i.e. that the pressure and energy density are dominated by
relativistic particles;
a steady state outflow; and that
the supernova core is the dominant source of gravity.

The specific net heating rate $q$ includes 
a number of processes. Heating
processes include (anti)neutrino absorption on nucleons, $\nu e$
scattering, and $\nu \bar{\nu}$ annihilation. The major cooling
processes are $e^{\pm}$ capture on free nucleons 
and $e^{\pm}$ annihilation.
 All but the last of these processes
depend on neutrino luminosities, energy moments of the neutrino
distribution functions, and geometric factors 
involving the maximum
neutrino deviation angle from the radial direction at a given 
radius. Expressions for these rates are given in QW. General
relativistic effects can be introduced through redshift factors
in the energy moments, and redshift and time dilation factors in 
luminosities (\cite{fuller1}); and by suitably 
altering the geometric factors to
account for curved neutrino trajectories (\cite{cardall1}).

Before proceeding to semi-analytic estimates, we briefly discuss
initial and boundary conditions. Treatment begins at the surface
of the supernova core, taken to be Schwarzschild coordinate 
radius $R$. As in QW, the initial entropy is taken to 
be $S_i =4$ 
($S > 4$ assures radiation dominated
conditions). The initial
temperature $T_i$ is obtained by equating the heating and cooling
rates at $R$. In QW neutrino absorption could be taken as the
sole heating process. This allowed a fairly simple expression
for $T_i$ to be obtained in terms of the 
$\bar{\nu}_e$ luminosity, 
the supernova core radius $R$, and the ratio of the second and 
first energy moments of the $\bar{\nu}_e$ spectrum 
($\epsilon_{\bar{\nu}_e}
\equiv \langle E_{\bar{\nu}_e}^2 \rangle / \langle 
E_{\bar{\nu}_e} \rangle$).  
For more compact cores, however, the other heating rates become
more important at the neutron star surface 
and it becomes impossible to obtain
an analytic expression for $T_i$ in terms of the 
above quantities. 
Therefore we choose the $\bar{\nu}_e$ luminosity 
to be $10^{51}$ erg s$^{-1}$, 
$M = 1.4 M_{\sun}$, and  $\epsilon_{\bar{\nu}_e} = 
20$ MeV, and numerically
solve for $T_i$ for various values of $2M/R$. 

Neutrino heating becomes negligible by
$T \approx 0.5$ MeV, where $\epsilon_{\rm flow}$ and $S$ reach 
their final values, after which they 
remain constant. These final values
will be denoted by a subscript $f$.  We assume
that a boundary is provided by the supernova shock at some large 
radius, where the temperature is $T_b$, and where the
flow asymptotes to small velocity. We then have 
from Eq. (\ref{eflow})
\begin{equation}
	{\ln\left(1+{{TS_f\over m_N}}\right)}
	- {1\over 2}{\ln\left(1-{v^2}\right)} +
	 {{1\over 2}\ln\left(1-{ 2 M\over r}\right)} \approx
	\epsilon_{\rm flow,f} \approx 
	\ln \left(1+{{T_b S_f\over m_N}}\right) \label{boundary}
\end{equation}
between $T=0.5$ MeV and $T_b$.

We now obtain an approximate expression for $S_f$. 
From Eq. (\ref{eflow}),
\begin{eqnarray}
\int_R^{r_f} {q \over (1 + TS/m_N)}{dr \over vy} &=&
	\epsilon_{{\rm flow},f} - \epsilon_{{\rm flow},i}, \\
& & \nonumber \\
{1 \over (1 + T_{\rm eff}S_{\rm eff} /m_N)} \int_R^{r_f} q \,
	{dr \over vy} &\approx& \ln \left(1 + 
	{T_b S_f \over m_N}\right) -
	{1\over 2}\ln\left(1 - {2M\over R}\right),
\end{eqnarray}
where $T_{\rm eff}$ and $S_{\rm eff}$ will be defined below.
From Eq. (\ref{entropy1}),
\begin{equation}
S_f \approx S_f - S_i = m_N \int_R^{r_f} {q\over T}{dr \over vy}.
\end{equation}
Combining these last two equations, taking 
$\ln \left(1 + {T_b S_f / m_N}\right) \approx {T_b S_f / m_N}$,
and using the definitions (QW) $S_{\rm eff} \approx S_f / 2$ and
$T_{\rm eff} \equiv  \left[\int (vy)^{-1} q\, dr\right] 
/ \left[\int (vyT)^{-1} q\, dr \right]$, 
we obtain
\begin{equation}
S_f \approx - {m_N \ln\left(1 - {2M/ R}\right) 
\over 2 T_{\rm eff} }
	\left[ 1 - {T_b \over T_{\rm eff} }- {1\over 4}
	\ln\left(1 - {2M\over R}\right) \right]^{-1}. 
	\label{entropy2}
\end{equation}
The quantity 
$T_{\rm eff}$ is an average temperature, weighted by 
the heating rate. In QW it is approximated as the temperature at
which the 
heating rate is a maximum. Using only neutrino absorption 
and $e^{\pm}$ 
annihilation, they obtain $T_{\rm eff} = 6^{-1/6}\, T_i$. 
As mentioned previously in connection with the determination of
the initial temperature $T_i$, addition of the other heating and
cooling processes makes analytic progess impossible.  To obtain
approximate numerical results we will
employ the 
expression  $T_{\rm eff} = 6^{-1/6}\, T_i$ as obtained by QW.
Figure 1 is a 
plot of S as a function of $2M/R$, obtained from Eq.
(\ref{entropy2}) with $T_b = 0.1$ MeV. 
The circle shows the numerical
result obtained in QW for their 
model 10B with post-Newtonian corrections.
The agreement is excellent.

Next we estimate $\dot M$. Baryon mass conservation in the 
Schwarzschild geometry gives 
$\dot M = 4\pi r^2 \rho_b vy = \rm{constant}$, 
where $\rho_b$ is the baryon mass density (\cite{cardall1}). 
As in QW, we evaluate
this at $r_{\rm eff} \approx R$ to obtain
$\dot M = 4\pi R^2\, \rho_{b, {\rm eff}}\, (vy)_{\rm eff}$.
The quantity $\rho_{b, {\rm eff}}$ can be derived from 
the entropy per baryon $S_{\rm eff} = (11\pi^2 /45)
(m_N T_{\rm eff}^3 / \rho_{b, {\rm eff}})$. To estimate
$(vy)_{\rm eff}$, we use Eq. (\ref{eflow}),
\begin{equation}
S_{\rm eff} \approx m_N \int_{T_i}^{T_{\rm eff}} {q \over vy}
	{dr \over dT}{dT\over T} \approx 
	{m_N \, q_{\rm eff} \over (vy)_{\rm eff}}
	\left| {dT \over dr}\right| ^{-1}_{\rm eff} 
	\ln \left({T_i \over 
	T_{\rm eff}}\right),
\end{equation}
which can be solved for $(vy)_{\rm eff}$.
The quantity $q_{\rm eff}$ is obtained by evaluating the net
heating rate $q$ at temperature $T_{\rm eff}$ and radius
$r_{\rm eff} \approx R$.  We follow QW and estimate the
temperature gradient by assuming approximate hydrostatic
equilibrium near radius $R$; the result is (\cite{cardall1}) 
\begin{equation}
{S\over m_N} \left| {dT\over dr}\right| \approx {(1 + TS/m_N)
	\over (1 - 2M/r)}{M \over r^2}.
\end{equation}
Now $\dot M$ can be
calculated using the previously obtained values of $S$.
A plot of $\dot M$ as a function of $2M/R$ is given
in Figure 2. (Note the kink at $R=3M$; below this radius,
there is a maximum angle of deviation from the radial direction
beyond which massless particles cannot escape to infinity.)

The dynamic expansion time scale 
$\tau_{\rm dyn}$ is defined to be
\begin{equation}
\tau_{\rm dyn} \equiv {1\over (vy)_f} \left|{1\over T}
	{dT\over dr}\right|_f^{-1},
\end{equation}
where the subscript $f$ means that the quantity is evaluated
at $T=0.5$ MeV. The dynamic expansion time scale is closely
related to the proper time a fluid element experiences
in going from $T_1=0.5$ MeV to, say $T_2=0.2$ MeV:
\begin{equation}
\int_{T_1}^{T_2} 
	d\tau_{\rm proper} = \int_{T_1}^{T_2} {dr \over vy} =
 \int_{T_1}^{T_2} {T \over vy}{dr\over dT} {dT\over T}
 \approx  {1\over (vy)_f} \left|{1\over T}
	{dT\over dr}\right|_f^{-1} \ln\left({0.5 
	\over 0.2} \right)
	\approx \tau_{\rm dyn}.
\end{equation}
In this case, we estimate the temperature scale height by
approximating the boundary condition in Eq. (\ref{boundary})
as $TS_f / m_N \approx M/r$. Then 
\begin{equation}
\left|{1\over T}{dT\over dr}\right|_f^{-1} \approx r_f \approx
	{M m_N \over S_f T_f}. 
\end{equation}
The quantity $(vy)_f$ can be obtained from the previously
computed quantities $\dot M$ and 
$S_f$. A plot of $\tau_{\rm dyn}$
as a function of $2M/R$ is given in Figure 3.

Neutron-to-seed ratios on the order of 100 or greater are
necessary for a successful $r$-process, and this can be achieved
under a variety of combinations of $S$, $\tau_{\rm dyn}$, and
$Y_e$ (Hoffman et al. 1996; 
Meyer et al. 1996; Meyer \& Brown 1997). 
We have here confirmed that general
relativistic
effects increase $S$ and reduce $\tau_{\rm dyn}$, 
as seen in selected
numerical calculations in QW. Both of these trends lead to a 
higher neutron-to-seed ratio.  On the other hand, 
Fuller \& Qian (1996) have shown
that general relativistic effects tend to increase $Y_e$ because
of the differential redshift of the $\nu_e$ and 
$\bar{\nu}_e$ emitted
from the supernova core. This differential 
redshift decreases the difference
in the average energies of the $\nu_e$ and 
$\bar{\nu}_e$ populations,
driving $Y_e$ larger and so 
closer to 0.5. This is probably unfavorable
for a high neutron-seed-ratio, 
since the general trend is that a higher
$Y_e$ requires a higher $S$ to obtain a given 
neutron-to-seed ratio.
The magnitude of the differential redshift effect 
on $Y_e$ is uncertain
because  of its dependence on unknown details of 
the nuclear equation
of state.

On the whole, however, general relativistic 
effects are likely to increase   
the neutron-to-seed ratio. For one thing, an 
increased $Y_e$ due to
differential neutrino redshifts is not always bad: 
examination of
Figure 10 and Table 5 of Hoffman et al. (1996) 
shows that the entropy 
requirements to obtain a given neutron-to-seed ratio 
actually become
{\em less} severe as $Y_e$ gets very close to 0.5. 
More important than
this, however, is the fact that salutory effects on 
$S$ and $\tau_{\rm dyn}$ can 
more than compensate for deleterious 
differential redshift effects
on $Y_e$.  For example, consider the case of 
$2M/R = 0.68$, for which our
semianalytic estimates give $\tau_{\rm dyn} \simeq 0.004$ s and
$S \simeq 255$.  In Table 5 of Hoffman et al. (1996), for 
``expansion time''
0.005 s (which corresponds to $\tau_{\rm dyn} \simeq 0.004$ s), 
an entropy of only 147 
is needed to obtain a neutron-to-seed ratio 
$\gtrsim 100$, even for the ``worst case'' value of $Y_e$. 

It should be pointed out that the case we just 
considered,  $2M/R = 0.68$,
(just) violates the causality 
limit $R > 1.52 \times 2M$ imposed on any
equation of state allowing a 
stable neutron star. However, we only 
consider this particular
case for comparison with 
Hoffman et al. (1997). Unfortunately, their Table 5
does not contain values of 
expansion time between 0.005 s and 0.025 s.
We fully expect that there are values of expansion time between
0.005 s and 0.025 s for which 
$2M/R$ does not violate the causality limit,
and for which $S$ is sufficiently high to provide a
neutron-to-seed ratio in 
excess of 100 even for the ``worst case'' $Y_e$.
Also, even beyond the 
causality limit, our calculations may give an idea
of what happens near an unstable supernova core that collapses
to a black hole at 
relatively late times. Of course, the time scale of collapse
may be fast enough to 
grossly violate our assumption of static conditions.
Reliable exploration of 
this case would require detailed numerical modeling
and good knowledge of the equation of state of nuclear matter.

There are other questions regarding the viability of a highly
relativistic supernova 
core as a suitable environment for $r$-process
nucleosynthesis.
While a small dynamic expansion time scale is 
conducive to a large neutron-to-seed
ratio, it may not allow 
enough neutron capture time to build up the
$r$-process elements from 
the seed nuclei. Some mechanism would have
to exist to slow down the flow during the neutron capture epoch
(see e.g. McLaughlin \& Fuller 1997); perhaps
the shock used in the boundary condition could serve this
purpose. Also, we have 
seen that a more compact core leads to a smaller
mass outflow rate.  One of the attractions of neutrino-heated 
supernovae ejecta as an 
$r$-process site is that 
the total mass loss, together with the estimated
Galactic supernova rate, roughly fits the observed amount of
Galactic $r$-process material (Meyer et al. 1992).
One would not want $\dot M$ to 
become so small that this agreement is
ruined. However, even for the 
extreme case of $2M/R = 0.68$ considered
above, our estimated 
$\dot M \simeq 1.3 \times 10^{-6} M_{\sun} s^{-1}$ may be 
large enough to be viable.

\acknowledgments
We wish to thank Y.-Z. Qian and J. R. Wilson for helpful
discussions. This work was supported 
by grants NSF PHY95-03384 and NASA NAG5-3062
at UCSD.

\newpage

\begin{figure}
\plotone{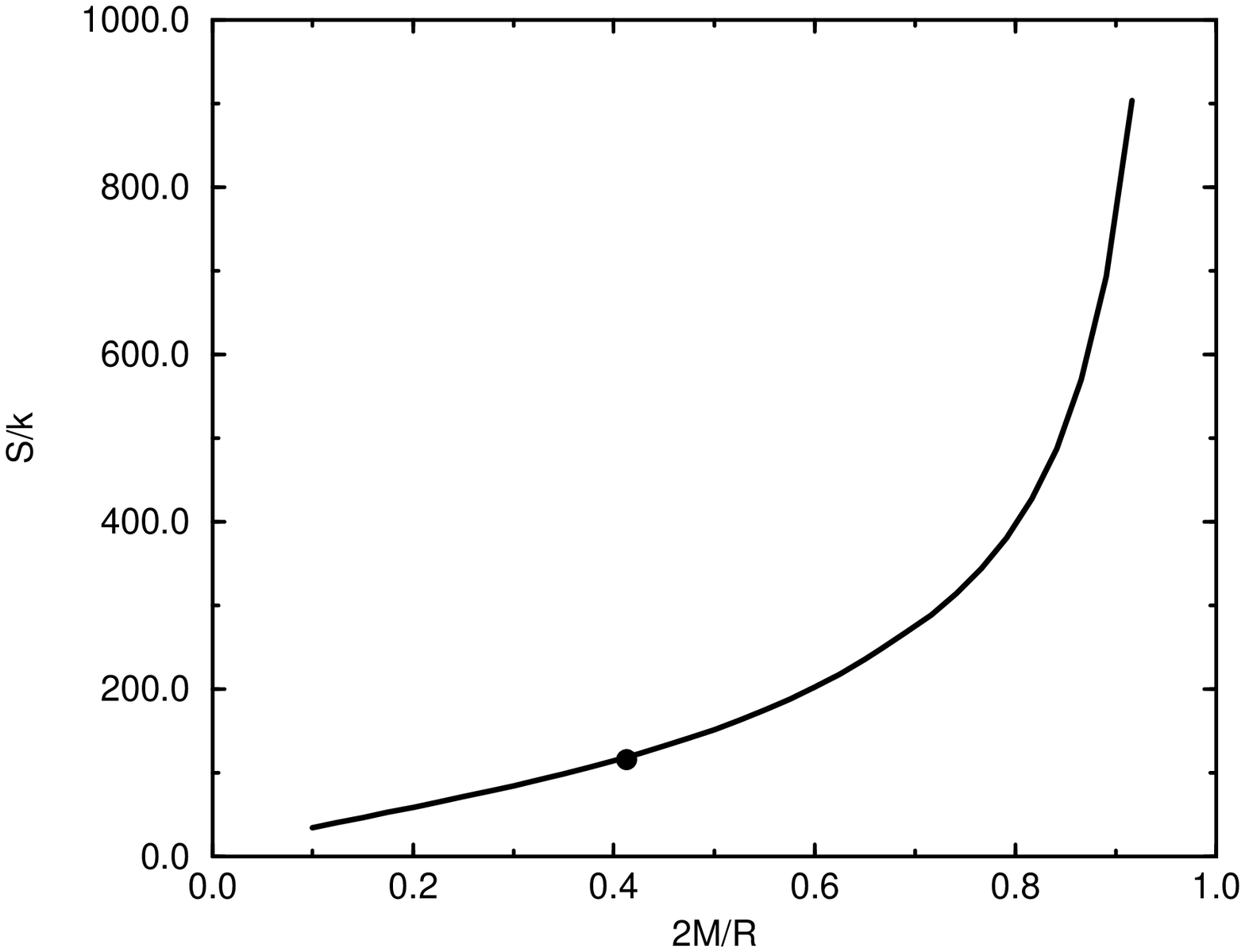}
\caption{The final entropy per 
	baryon in units of Boltmann's
	constant, as a function 
	of the supernova core Schwarzschild radius
	divided by the core radius. 
	The circle is from the Qian \& Woosley (1996) 
	numerical calculation of 
	model 10B with post-Newtonian corrections.}
\end{figure}

\begin{figure}
\plotone{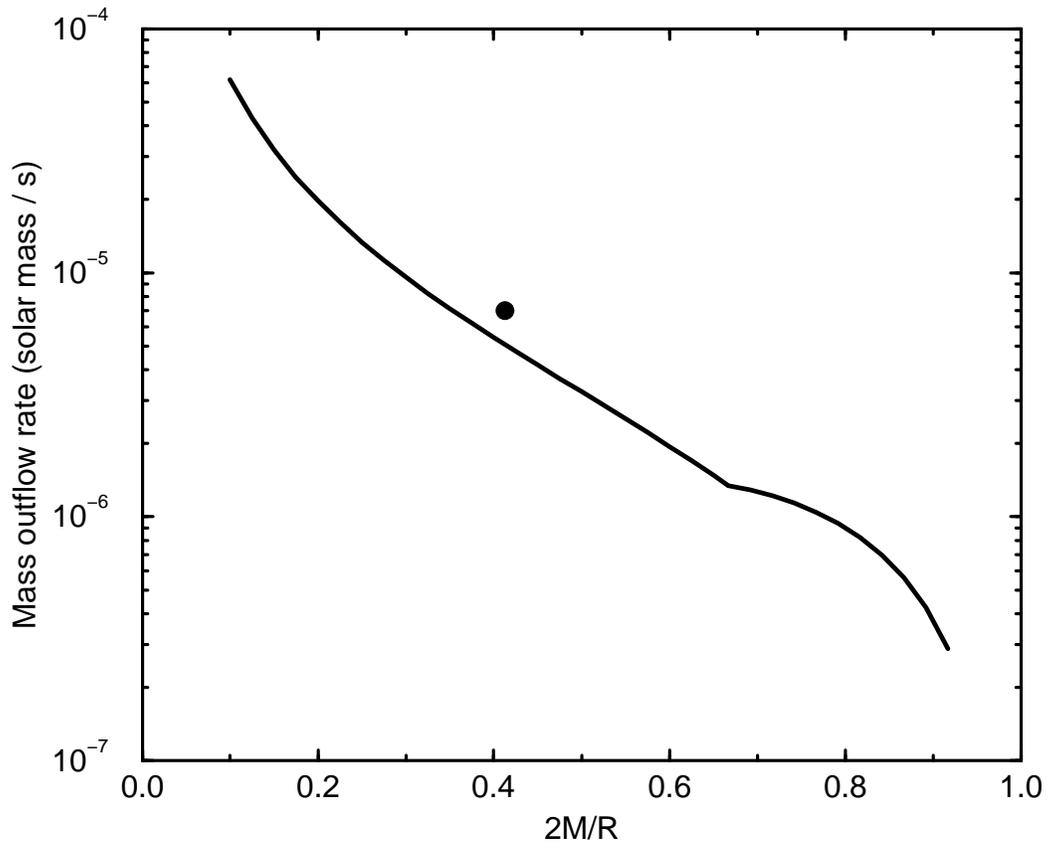}
\caption{The mass outflow rate as a 
	function of the supernova core Schwarzschild radius
	divided by the core radius. 
	The circle is from the Qian \& Woosley (1996) 
	numerical calculation of 
	model 10B with post-Newtonian corrections.}
\end{figure}

\begin{figure}
\plotone{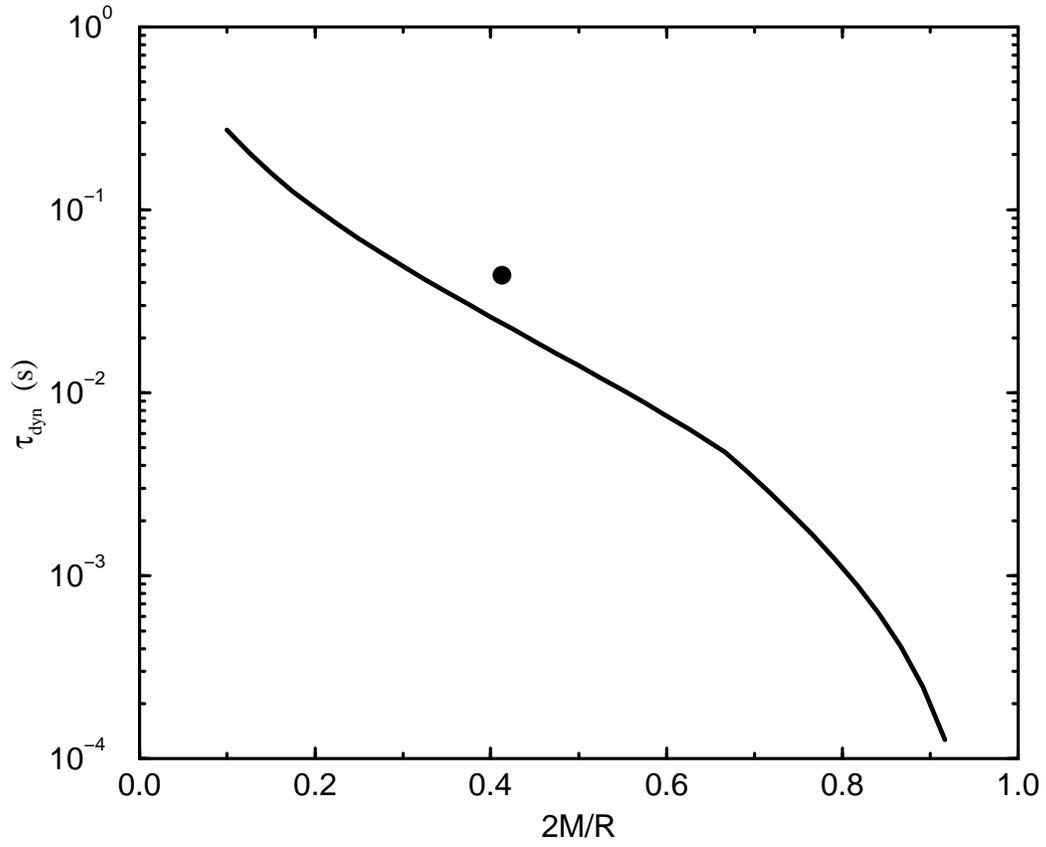}
\caption{The dynamic expansion time scale 
	as a function of the supernova core Schwarzschild radius
	divided by the core radius. 
	The circle is from the Qian \& Woosley (1996) 
	numerical calculation of 
	model 10B with post-Newtonian corrections.}
\end{figure}

\end{document}